\documentclass[aps,prl,twocolumn,showpacs]{revtex4}
\usepackage{graphicx}
\usepackage{amssymb}
\newcommand{ \be}{\begin{equation}}
\newcommand{ \ee}{\end{equation}}
\newcommand{\beq}{\begin{eqnarray}}
\newcommand{\eeq}{\end{eqnarray}}
\newcommand{\bem}{\begin{pmatrix}}
\newcommand{\eem}{\end{pmatrix}}
\newcommand{\bmx}{\begin{array}}
\newcommand{\emx}{\end{array}}

\begin{document}

\title{Freezing and melting of 3D complex plasma structures under microgravity conditions driven by neutral gas pressure manipulation}

\author{S. A. Khrapak,$^{1,2}$ B. A. Klumov,$^{1,2}$ P. Huber,$^1$ V. I. Molotkov,$^2$ A. M. Lipaev,$^2$ V. N. Naumkin,$^2$ H. M. Thomas,$^1$ A. V. Ivlev,$^1$ G. E. Morfill,$^1$ O. F. Petrov,$^2$ V. E. Fortov,$^2$ Yu. Malentschenko,$^3$ and S. Volkov$^3$}

\affiliation{$^1$Max-Planck-Institut f\"ur extraterrestrische Physik, D-85741 Garching,
Germany \\$^2$Joint Institute for High Temperatures, 125412 Moscow, Russia \\$^3$Yuri Gagarin Cosmonaut Training Centre, 141160 Star City, Russia}

\date{\today}

\begin{abstract}
Freezing and melting of large three-dimensional complex plasmas under microgravity conditions is investigated. The neutral gas pressure is used as a control parameter to trigger the phase changes: Complex plasma freezes (melts) by decreasing (increasing) the pressure. Evolution of complex plasma structural properties upon pressure variation is studied. Theoretical estimates allow us to identify main factors responsible for the observed behavior.
\end{abstract}

\maketitle

Complex (dusty) plasmas -- systems consisting of highly charged micron-size particles in a neutralizing plasma background -- exhibit an extremely rich variety of interesting phenomena ~\cite{Revs}. Amongst these the transitions between fluid and solid phases (freezing and melting) are of particular interest~\cite{Revs,Crystal,Morfill_Nat,melt_freez}.
This is largely a consequence of the fact that high temporal and spatial resolution allows us to investigate these phase changes along with various related phenomena at the individual particle level~\cite{Revs,Crystal,Morfill_Nat,melt_freez,Milenko,Knapek,Couedel}.

In this Letter we report on experimental studies of the solid-liquid phase changes in large 3D complex plasmas under microgravity conditions driven by manipulating neutral gas pressure. It is observed that the system of charged particles exhibits freezing (melting) upon decreasing (increasing) the pressure, in contrast to the situation in ground-based experiments where plasma crystals melt upon {\it reducing} the pressure. This can illustrate important differences between generic (e.g. similar to colloidal suspensions) and plasma-specific mechanisms of phase transitions in complex plasmas.

Experiments are performed in the PK-3 Plus laboratory operating onboard the International Space Station~\cite{PK3+}. The heart of this laboratory is a parallel-plate radio-frequency (rf) discharge operating at a frequency of 13.56 MHz. Discharge can operate in argon, neon or their mixture in a wide range of pressures, rf-amplitudes and rf-powers. Complex plasmas are formed by injecting monodisperse micron-size particles into the discharge. The particles are illuminated by a diode laser and the scattered light is recorded by CCD cameras. For a comprehensive review of the PK-3 Plus project see Ref.~\cite{PK3+}.

\begin{figure}
\includegraphics[width=8.0cm]{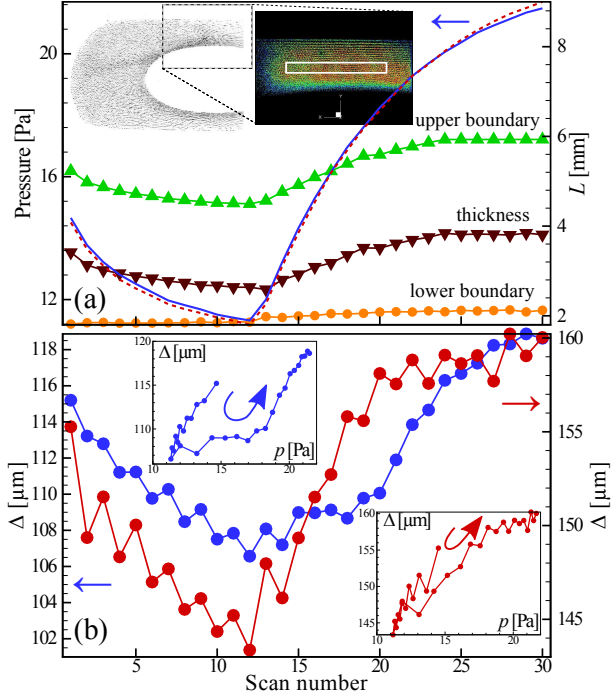}
\caption{(Color) (a) Positions of the upper and lower boundaries of the particle cloud and its thickness in the vertical direction vs. the scan number for the experimental run with small particles. The corresponding values of pressure are shown for both runs [red dashed (blue solid) curves for small (big) particles]. Left inset shows side view of the particle cloud (inverted colors), right inset corresponds to the FoV of the high resolution camera [particles are color-coded to see solid-like (red) and liquid-like (blue) domains]. Rectangle marks the part of the cloud used for the detailed structural analysis (rectangular box $7\times 0.7 \times 4.5$ mm$^3$); (b) Interparticle distance vs. scan number in two experimental runs. Insets show the dependence $\Delta(p)$ demonstrating some hysteresis.}
\label{boundaries}
\end{figure}

The experiments described here are carried out in argon at a low rf-power ($\sim 0.5$ W). We use two different sorts of particles in the two distinct experimental runs: SiO$_2$ spheres with a diameter $2a =1.55$ $\mu$m and Melamine-Formaldehyde spheres with $2a =2.55$ $\mu$µm. The experimental procedure, identical in these two runs, is as follows: When the particles form a stable cloud in the bulk plasma, the solenoid valve to the vacuum pump is opened, which results in a slow decrease of the gas pressure $p$. Then, the valve is closed and $p$ slowly increases due to the gas streaming in. (Neutral flow has negligible direct effect on the particles). During the pressure manipulation ($\simeq 6$ minutes in total), the structure of the particle cloud is observed. The observations cover the pressure range from $\simeq 15$ Pa, down to the lowest pressure of $\simeq 11$ Pa and then up to $\simeq 21$ Pa (see Fig.~\ref{boundaries}a). In order to get three-dimensional particle coordinates, 30 scans are performed. Scanning is implemented by simultaneously moving laser and cameras in the direction perpendicular to the field of view with the velocity 0.6 mm/s. Each scan takes $\simeq 8$ s, resulting in the scanning depth of $\simeq 4.8$ mm; the interval between consecutive scans is $\simeq 4$ s. The particle positions are then identified by tomographic reconstruction of the 3D-pictures taken with the high resolution camera, which observes a region of about $8 \times 6$ mm$^2$ slightly above the discharge center.

Let us first analyze the global reaction of the particle cloud on pressure manipulation. Figure \ref{boundaries}a shows the positions of the upper and lower cloud boundaries and the cloud thickness as a function of the scan number (time). It is evident that the position of the upper boundary is strongly correlated with pressure: It moves downwards (upwards) with the decrease (increase) of $p$. This has a clear physical explanation. Particles cannot penetrate in the region of strong electric field (sheath) established near the upper electrode. The position of the upper cloud boundary is thus set by the sheath edge. The sheath thickness is roughly proportional to the electron Debye radius $\lambda_{{\rm D}e}$ which exhibits the following approximate scaling $\lambda_{{\rm D}e}\propto n_e^{-1/2}\propto p^{-1/2}$. This implies that upon a decrease (increase) of $p$, the particles are pushed farther (closer) to the electrode, in full agreement with the observations. The lower cloud boundary, associated with the presence of the particle-free region (void) in the central area of the discharge \cite{Morfill99,LipaevPRL}, shows less systematic behavior. Its position does not change when $p$ decreases, but then moves upwards monotonously when $p$ increases. However, the displacement amplitude is relatively small. As a result, the thickness of the cloud exhibits pronounced decrease (increase) when $p$ decreases (increases), see Fig.~\ref{boundaries}a. Thus, the particle component becomes compressed by reducing the pressure. The resulting dependence of the mean interparticle distance (in the part of the particle cloud subject to detailed analysis) on the scan number/pressure is shown in Fig.~\ref{boundaries}b. The interparticle distance $\Delta$ is clearly correlated with $p$, although some hysteresis (more pronounced for small particles) is evident from insets in Fig.~\ref{boundaries}b.

\begin{figure}
\includegraphics[width=8.0cm]{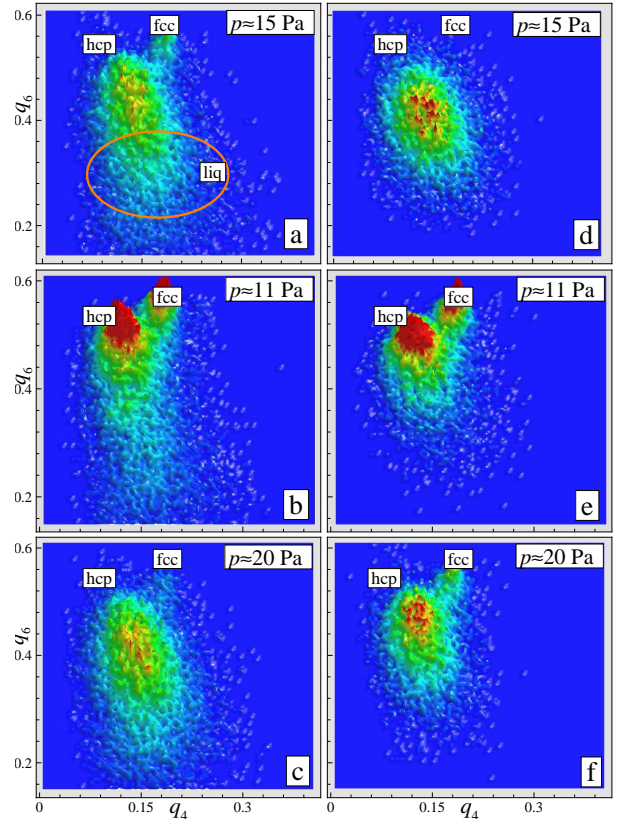}
\caption{(Color) Variation of the structural properties with pressure as reflected by particle distributions on the plane of rotational invariants ($q_4, q_6$) for the system of small (left panel) and big (right panel) particles. The rotational invariants for perfect hcp and fcc lattices and liquid-like domain [sketched in (a)] are also indicated. For discussion see text.}
\label{panel}
\end{figure}

For the detailed analysis of the structural properties during the pressure manipulation, a part of the cloud sketched in Fig.~\ref{boundaries} has been chosen. It contains of the order of $10^4$ particles. To determine the local order of particles we use the bond order parameter method \cite{stein}. In this method, the local rotational invariants $q_i,~w_i$ \cite{stein,3Da,3Db}
for each particle are calculated and compared with those for ideal lattice types. Here, we are specifically interested in identifying face-centered cubic (fcc), body-centered cubic (bcc), and hexagonal close-packed (hcp) lattice types and, therefore use the invariants $q_4$, $q_6$, $w_6$ calculated using 12 and 8 nearest neighbors. Figure~\ref{panel} shows representative examples of particle distributions on the plane ($q_4$, $q_6$). Initially, both systems of smal (Fig.~\ref{panel}a) and big (Fig.~\ref{panel}d) particles reveal weakly ordered liquid-like phase. Upon a decrease in $p$, the particles tend to form more ordered structures, and eventually both systems freeze. The distributions corresponding to the minimum pressure are shown in Figs.~\ref{panel}b, e. Clear crystalline structures which are dominated by the hcp and fcc lattices are observed. Subsequent increase in pressure destroys the particle ordering. Figures \ref{panel} c,f demonstrate the final states of the systems.

Scan by scan analysis of the complex plasma structural composition reveals the following main properties: (i) A decrease (increase) in $p$ enhances (suppresses) ordering of the particles; (ii) Maximum number of particles in the crystalline state corresponds to the minimal pressure; (iii) Crystalline phase is mostly composed of hcp- and fcc-like particles with only a small portion of bcc-like clusters; (iv) The system of small particles exhibits melting with increasing $p$, whilst the system of big particles approaches the melting transition, but does not fully melt; (v) Premelting stage (disappearance of fcc- and bcc-like particles \cite{3Da}) is observed for small particles, but is less pronounced in the system of big particles.

\begin{figure}
\includegraphics[width=7.3cm]{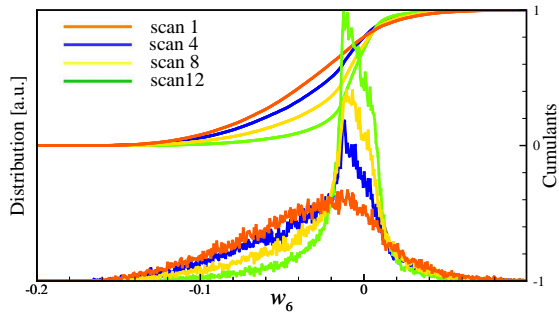}
\caption{(Color online) Distributions of small particles over the rotational invariant $w_6$ for different scan numbers (indicated in the figure). The corresponding cumulants $\hat W_6(x)$ are plotted in the upper part of the figure. The position of the half-height of $\hat W_6(x)$ is used as a melting indicator (see text).}
\label{w_6}
\end{figure}

\begin{figure}
\includegraphics [width=7.3cm] {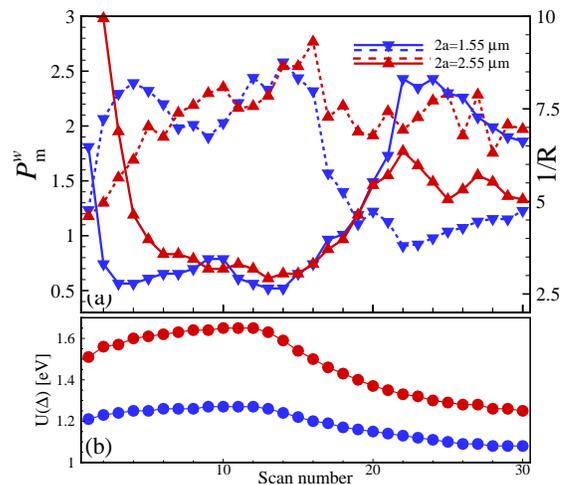}
\caption{(Color online) (a) Values of the melting (freezing) indicator $P_m^{w}$ ($R^{-1}$) vs. the scan number. Solid curves correspond to $P_m^{w}$, dashed curves to $R^{-1}$. System of small (big) particles is denoted by blue (red) color. (b) Estimated interparticle interaction energy for each scan.
}
\label{op}
\end{figure}

It has been shown recently  that cumulative distribution function $\hat W_6(x) \equiv \int_{-\infty}^x n(w_6)dw_6$ is very sensitive to phase state of various systems \cite{melt_freez,PRB}. Here
$n(w_6)$ is the distribution, normalized to unity, of particles over the rotational invariant $w_6$. Figure~\ref{w_6} shows $n(w_6)$ and the corresponding $\hat W_6(w_6)$ for four different scans corresponding to different states of the system of small particles. An appropriate melting indicator can be defined as $P_m^{w} \equiv w_6^{\rm hh}/w_6^{\rm fcc}$, where $w_6^{\rm hh}$ is the position of the half-height of $\hat W_6(w_6)$ [$\hat W_6(w_6^{\rm hh}) =1/2$] and $w_6^{\rm fcc}=-1.3161\times 10^{-2}$ is the value of $w_6^{\rm hh}$  for the fcc lattice.
Melting occurs when $P_m^{w} \gtrsim 1.3$. There also exists a freezing criterion~\cite{rdf} based on the properties of the pair correlation function $g(r)$. It states that near freezing, the ratio $R=g^{\rm min}/g^{\rm max}$, where  $g^{\rm max}$ ($g^{\rm min}$) is the value of $g(r)$ at the first maximum (first nonzero minimum), is a constant, $R\simeq 0.2$. This rule describes fairly well freezing of the Lennard-Jones fluid, but is not universal: $R$ can vary considerably for repulsive potentials of different softness~\cite{Agrawal}. However, softness of the interactions in the considered complex plasmas is expected to fall in the range, where $R$ is nearly constant. Fig.~\ref{op}a shows melting and freezing indicators $P_m^{w}$ and  $R$ for different scans. Both criteria clearly indicate that the structures are mostly ordered near the lowest pressure and identify freezing and melting at similar conditions [except system of big particles, which does not melt (is near melting) according to the value of $R$ ($P_m^{w}$)].

We interpret the observed fluid-solid phase changes to the variation in the electrical repulsion between highly charged particles. Manipulating the gas pressure experimentally, changes various complex plasma parameters and modifies the strength of the repulsion. When the electrical coupling reach (or drops below) certain level, freezing (or melting) occurs. To verify this scenario we need to estimate relevant plasma parameters.

We use the results from siglo-2D simulations \cite{PK3+} to estimate plasma parameters in the absence of particles. In the considered regime ($p \sim 10-25$ Pa, rf-amplitude $\sim 15$ V) the central plasma density is linear on $p$ and can be, with a reasonable accuracy, described as $n_0\simeq (1.20+0.11p)\times 10^8$, where $n_0$ is in cm$^{-3}$ and $p$ in Pa. The electron temperature exhibits almost no dependence on $p$, $T_e\simeq 3.8$ eV. Ions and neutrals are at room temperature $T_{i,n}\sim 0.03$ eV. When particles are injected into the discharge they inevitably modify plasma parameters. In the following we assume that {\it inside the particle cloud} the electron temperature remains unaffected, while the electron and ion densities are modified to keep quasineutrality, $n_e+|Q/e|n_p\simeq n_i$, where $Q$ is the particle charge. Furthermore, we assume that $n_e$ remains close to the particle-free value $n_0$, while $n_i$ somewhat increases in response to perturbations from the particle component. This approximation is in reasonable agreement with numerical simulation results~\cite{Land,Goedheer} regarding plasma parameters {\it inside the particle cloud} (note that {\it inside the void region} in the center of the discharge, the plasma density and electron temperature can be considerably higher than those in the particle-free discharge \cite{Land,Goedheer}). Using these assumption we can calculate the dependence $Q(p,\Delta)$ employing the collision enhanced collection (CEC) approximation \cite{CEC} for the ion flux and the OML model for the electron flux to the particle. The flux balance condition yields $v_{T_i}(1+zP)(1+z\tau+0.1\frac{z^2\tau^2}{\sqrt{1+zP}}\frac{\lambda_0}{\ell_i})=v_{T_e}\exp(-z)$, where $z=|Q|e/aT_e$ is the reduced particle charge, $P=(aT_e/e^2)(n_p/n_0)$ is the scaled ratio of particle-to-plasma densities (the so-called Havnes parameter), $\tau=T_e/T_i$ is the electron-to-ion temperature ratio, $v_{T_e(i)}=\sqrt{T_{e(i)}/m_{e(i)}}$ is the electron (ion) thermal velocity, $\lambda_0=\sqrt{T_i/4\pi e^2n_0}$ is the unperturbed ion Debye radius, and $\ell_i\simeq T_n/p\sigma_{in} $ is the ion mean free path ($\sigma_{in}\simeq 2\times 10^{-14}$ cm$^2$ is the effective ion-neutral collisions cross section in argon). The interaction potential is assumed to be of Yukawa form, the interaction energy between neighboring particles is $U=(Q^2/\Delta)\exp(-\Delta/\lambda)$, where $\lambda=\lambda_0/\sqrt{1+zP}$ approximates the plasma screening length. We neglect long-range inverse-power-law corrections to the Yukawa potential~\cite{KKM} since the ratio $\kappa= \Delta/\lambda$ is not large enough in the present experiments ($\kappa \simeq 3$ and $4$ for small and big particles, respectively).

Using the obtained values of $\Delta$, $Q$, and $\lambda$ we can estimate variations in $U$. The results are shown in Fig. \ref{op}b. The energy, as a function of $p$, exhibits maximum at $p\simeq 11$ Pa. The observed trends are in good qualitative agreement with the results of the structural analysis. From the observation that at high pressures the system of big particles is near melting  (Fig.~\ref{op}a) we can roughly estimate the particle temperature, $T_p\lesssim 5T_n\simeq 0.15$ eV, by using $U/T_p\simeq 7.5$ at melting of the Yukawa system with $\kappa = 4$ \cite{KM_2009}. This estimate is sensitive to the input values of $T_e$ and $n_0$. To reach equilibrium ($T_p\simeq T_n$) we need to decrease $T_e$ or increase $n_0$ by approximately a factor of 7. Unfortunately,  $T_p$ is unknown here since the particle dynamics is not resolved. Finally, our best estimate of the relaxation time (defined as a time needed for a particle to diffuse over the distance $\Delta$) yields $\sim 1$ s ($\sim 2$ s) for the system of small (big) particles in the {\it fluid state near freezing}. This is considerably shorter than the characteristic time associated with external (pressure) perturbations. This implies our complex plasmas quickly reach equilibrium static configuration, i.e., they remain in quasistatic equilibrium, supporting our interpretation.

To conclude, let us discuss some specific properties of the phase changes observed. To identify the main mechanism responsible for the structural changes let us define the relative change of a parameter $f$ between the cases of highest (30th scan) and lowest (12th scan) pressures as $\delta f=(f_{30}-f_{12})/f_{12}$. Normally, one would expect a decrease in $Q$, an increase in $n_0$, and therefore a decrease in $\lambda$ when $p$ increases. However, the coupling between $n_e$, $n_i$, and $n_p$ and the corresponding charge reduction in dense particle clouds (which is to some extent similar to a reduction of colloidal charge when increasing colloidal volume fraction~\cite{Royall}) interferes. As a result, in the regime investigated, some increase in $\Delta$ with $p$ [$\delta \Delta \simeq 0.11 (0.12)$] is accompanied by almost constant particle charge [$\delta z \simeq 0.01 (0.00)$] and slight {\it increase} in $\lambda$ [$\delta \lambda\simeq 0.08 (0.07)$] for the system of small (big) particles, respectively. Thus, an increase in $\Delta$ (i.e. decrease in the particle density) is the main factor responsible for the melting. On the other hand, reducing pressure compresses the particle system and it freezes. This behavior is opposite to the conventional procedure of melting flat plasma crystals by {\it reducing} the pressure in ground-based experiments~\cite{Morfill_Nat,Melzer}. The difference is not a consequence of the essentially 2D character of crystals investigated on Earth, but is rather due to the presence of strong electric fields (and, therefore, strong ion flows) required to balance the force of gravity. There are effective mechanisms of converting the energy associated with ion flows into the kinetic energy of the particles. Known scenarios include ion-particle two-stream instability~\cite{Glenn}, nonreciprocity of the interaction due to asymmetric character of the screening cloud around the particles (``plasma wakes'')~\cite{Couedel,Schweigert}, particle charge variations~\cite{Ivlev}. All these scenarios lead to an abrupt increase of the particle kinetic energy at pressures below certain threshold value, causing crystal melting. The process of melting studied here is quite different from these plasma-specific mechanisms, but have much more in common with those in conventional molecular and soft matter (e.g. colloidal) systems.

This work was supported by DLR under Grant 50WP0203.

\end{document}